\documentclass[a4paper,12pt]{article}
\def\tev{{\rm TeV}}
\def\gev{{\rm GeV}}
\def\ten{\textbf{10}}
\def\five{\textbf{5}}

\def\etal{{\it et al.}}
\def\ie{{\it i.e.}}
\newcommand{\beq}{\begin{equation}}
\newcommand{\eeq}{\end{equation}}
\newcommand{\bea}{\begin{eqnarray}}
\newcommand{\eea}{\end{eqnarray}}
\newcommand{\bsub}{\begin{subequations}}
\newcommand{\esub}{\end{subequations} \noindent}
\newcommand{\clean}{\setcounter{equation}{0}}
\renewcommand{\theequation}{\thesection.\arabic{equation}}
\renewcommand{\labelenumi}{(\roman{enumi})}
\def\MPLA#1#2#3{Mod. Phys. Lett. {\bf A#1} (#2) #3}
\def\PRD#1#2#3{Phys. Rev. {\bf D#1} (#2) #3}

\def\NPB#1#2#3{Nucl. Phys. {\bf B#1} (#2) #3}
\def\PTP#1#2#3{Prog. Theor. Phys. {\bf #1} (#2) #3}

\def\PLB#1#2#3{Phys. Lett. {\bf B#1} (#2) #3}
\def\PRL#1#2#3{Phys. Rev. Lett. {\bf #1} (#2) #3}

\makeatletter
\newtoks\@stequation

\def\subequations{\refstepcounter{equation}%
  \edef\@savedequation{\the\c@equation}%
  \@stequation=\expandafter{\theequation}
  \edef\@savedtheequation{\the\@stequation}
  \edef\oldtheequation{\theequation}%
  \setcounter{equation}{0}%
  \def\theequation{\oldtheequation\alph{equation}}}

\def\endsubequations{%
  \ifnum\c@equation < 2 \@warning{Only \the\c@equation\space subequation
    used in equation \@savedequation}\fi
  \setcounter{equation}{\@savedequation}%
  \@stequation=\expandafter{\@savedtheequation}%
  \edef\theequation{\the\@stequation}%
  \global\@ignoretrue}

\def\eqnarray{\stepcounter{equation}\let\@currentlabel\theequation
\global\@eqnswtrue\m@th
\global\@eqcnt\z@\tabskip\@centering\let\\\@eqncr
$$\halign to\displaywidth\bgroup\@eqnsel\hskip\@centering
     $\displaystyle\tabskip\z@{##}$&\global\@eqcnt\@ne
      \hfil$\;{##}\;$\hfil
     &\global\@eqcnt\tw@ $\displaystyle\tabskip\z@{##}$\hfil
   \tabskip\@centering&\llap{##}\tabskip\z@\cr}

\makeatother

\begin{document}
\thispagestyle{empty}
\vspace*{-15mm}
\baselineskip 10pt
\begin{flushright}
\begin{tabular}{l}
{\bf OCHA-PP-178}\\
{\bf hep-ph/0105287}
\end{tabular}
\end{flushright}
\baselineskip 24pt 
\vglue 10mm 
\begin{center}
{\Large\bf
Mixing Matrix of Quarks Having Natural Twisting in $E_6$
Grand Unified Model
} 
\\
\vspace{15mm}

\baselineskip 18pt 
\def\thefootnote{\fnsymbol{footnote}}
\setcounter{footnote}{0}

{\bf 
Midori Obara$^{a)}$,Gi-Chol Cho$^{b)}$, \\
Makiko Nagashima$^{a)}$ 
and Akio Sugamoto$^{b)}$
}
\vspace{5mm}

{\it $^a$Graduate School of Humanities and Sciences,}  \\
{\it Ochanomizu University, Tokyo 112-8610, Japan}  \\
\vspace{2mm}

{\it 
$^b$Department of Physics,
Ochanomizu University, \\
Tokyo 112-8610, Japan} 
\vspace{10mm}
\end{center}
\begin{center}
{\bf Abstract}\\[7mm]
\begin{minipage}{14cm}
\baselineskip 16pt
\noindent
Mass matrix of quarks is studied in the Supersymmetric 
$E_6$ Grand Unified Theory (GUT). 
The fundamental representation $\textbf{27}$ in $E_6$ 
which corresponds to one generation contains two sets of 
$\textbf{5}^{*}$'s in SU(5), so that there are six flavors 
of lepton doublets and right-handed down-quark triplets. 
It is known that the twisting (interchange) among the 
$\textbf{5}^{*}$ representations may reproduce 
the observed quark and lepton mixing matrices. 
If $E_6$ is directly broken down to 
SU(3)$_C \times$SU(2)$_L \times$U(1)$_{Y}$ and two extra 
U(1)'s, 
the extra 3 sets of down-type quarks do not necessarily 
decouple from the quark mass matrix, under the appropriate 
choice of the U(1) flavor charges on the quark and Higgs fields. 
Then, by diagonalizing the 6 $\times$ 6 down-quark mass matrix, 
we find a certain set of parameters, in which the twisting 
between a pair of the right-handed down-quarks occurs naturally, 
reproducing the reasonable values for the $d$-, $s$- and 
$b$-quark masses,  
and for $V_{ud}$, $V_{us}$, $V_{cd}$, $V_{cs}$ and $V_{tb}$ of 
the CKM matrix elements.
As a by-product, one vector-like down-quark appears at the 
experimentally accessible TeV scale. 
\end{minipage}
\end{center}
\newpage
\baselineskip 18pt 
\section{Introduction} 

The discovery of the neutrino masses and large mixing 
angles at Super-Kamiokande \cite{SK} triggers the burst of 
investigations on the masses and mixing matrices of quarks 
and leptons. 
From a view of the grand unified theory (GUT), one of 
the most important tasks is to understand the different 
structure of the flavor mixing matrix between quarks 
and leptons. 
It is known that the quark mixing, which is described 
by the  Cabibbo-Kobayashi-Maskawa (CKM) matrix $V$ \cite{CKM}, 
is small, \ie, $V_{12} \approx \lambda \approx 0.22$ 
(Cabibbo angle), $V_{23} = \mathcal{O}(\lambda^{2})$, 
and $V_{13} = \mathcal{O}(\lambda^{3})$. 
On the other hand, the neutrino oscillation experiments 
tell us that the lepton flavor mixing which is expressed 
by the Maki-Nakagawa-Sakata (MNS) matrix $U$ \cite{MNS} 
may be large, probably being ``doubly maximal'': 
$U_{12} \approx U_{23} \approx 1/\sqrt2$ \cite{bi-max}. 

Bando and Kugo have proposed an idea which is possible to 
explain the flavor 
mixing matrices based on the Supersymmetric (SUSY) $E_6$ GUT, 
in which one family of quarks and leptons belongs to 
a fundamental representation $\textbf{27}$ of 
$E_6$ \cite{Bando-Kugo}.  
The $\textbf{27}$ representation is decomposed in terms 
of SO(10) and SU(5) as follows: 
\bea
\begin{array}{cccccccl}
\textbf{27} &=& \textbf{16} &+& \textbf{10} &+& \textbf{1}, 
                                  & ~~~~~[\mbox{SO(10)}] \\
            &=& (\textbf{10} + \textbf{5}^* + \textbf{1}) 
                &+& (\textbf{5} + \textbf{5}^*) 
                 &+& \textbf{1}. & ~~~~~[\mbox{SU(5)}]
\end{array}
\label{eq:decomposition}
\eea
As is well known, $\ten$ of SU(5) includes the left-handed 
down-type quarks $d_{L}$ as well as the left- and right-handed 
up-type quarks $u_{L}$ and $u_{R}^{c}$, while the left-handed 
leptons and the right-handed down-type quarks are usually 
assigned to $\five^*$. 
It is interesting that, in $E_6$ GUT, there are two sets of 
$\five^*$ --- $(\textbf{16},\five^*)$ and $(\ten,\five^*)$, 
where \textbf{16} and \textbf{10} denote the representations 
in SO(10). 
The representation $(\textbf{16},\five^*)$ is ``usually'' 
identified as the $\five^*$ representation in the ordinary 
SU(5) GUT, which consists of $(d_{R}^{c}, e_{L}, \nu_{L})$. 
On the other hand, the representation $(\ten,\five^*)$, which 
consists of $(D_{R}^{c}, E_{L}, N_{L})$, is ``usually'' 
expected to decouple from the SU(5) GUT since it has a gauge 
invariant mass term associated with the  $(\ten,\five)$ 
representation, $(D_{R}, E_{L}^{c}, N_{L}^{c})$. 
But, from the viewpoint of the Standard Model (SM), 
both $d_R^c$ in $(\textbf{16},\five^*)$ and $D_R^c$ 
in $(\ten,\five^*)$ could be candidates of the light 
down-type quarks because they have the same quantum number 
under the SM gauge group, 
SU(3)$_C \times$SU(2)$_L \times$U(1)$_Y$. 
Then, there is a freedom to interchange (called ``twisting'')  
between two $\textbf{5}^{*}$ representations if 
$(D_R^c,E_L, N_L)$ is not much heavier than 
$(d_R^c, e_L,\nu_L)$. 
The consequence of the twisting in the 3rd generation 
has been studied in ref. \cite{Bando-Kugo}. 

The flavor mixing matrices of quarks and leptons are 
defined from the charged current interactions in their 
mass eigenstates. 
Then, the twisting between $d_R^c$ and $D_R^c$ does not 
affect the quark flavor mixing because they do not couple 
to the $W$ boson. 
However, since the leptons embedded in $\five^*$ are 
left-handed, the twisting between $(\tau_L, \nu_{\tau L})$ 
and $(E_L, N_L)_3$ affects the charged current interactions. 
Then, the mixing matrix of leptons, especially the related 
part with the 3rd generation, can be drastically changed.  
As a consequence, one of the double maximal mixings, 
$U_{23}$, is naturally derived~\cite{Bando-Kugo}. 
The other interesting twisting mechanisms have been proposed 
some time ago~\cite{Yana}.

In ref.~\cite{Bando-Kugo}, the Froggatt-Nielsen 
mechanism~\cite{Frog} is used to give the hierarchical 
structure of the down-quark mass matrix. 
As pointed out by Froggatt and Nielsen~\cite{Frog}, 
the renormalization group equations of the Yukawa couplings 
for up- and down-quarks do not differ so much. 
Hence, it may be difficult to explain the hierarchical 
structure of the mass matrices in this way unless using 
the extra U(1)$_F$ flavor symmetry. 
From the U(1)$_F$ invariance, for example, the mass term 
of up-quarks takes the following form:
\begin{equation}
-{\cal L}_{mass}= y_{ij} \overline{\psi}_{i}\psi_{j}
                 H \left( \frac{\Theta}{M_{P}} 
                 \right)^{f_{i}+f_{j}+x},
\end{equation}
where $\psi_{i}$ and $H$ are the quark and Higgs fields, 
respectively. 
The indices $i$ and $j$ denote the generation, and 
$y_{ij}$ is the Yukawa coupling whose magnitude 
is assumed to be order one. 
The field $\Theta$ is an another Higgs field which is 
responsible for the U(1)$_F$ symmetry breaking, and $M_P$ 
is the Plank mass. 
We fix the U(1)$_F$ charge of the quark field $\psi_i$, which 
is denoted by $f_i$ as follows: 
\bea 
(f_1,f_2,f_3) =(3,2,0), 
\label{eq:u1charge}
\eea
while those of $H$ and $\Theta$ are taken as $x$ and $-1$, 
respectively~\cite{Bando-Kugo}. 
If the U(1)$_F$ symmetry breaking scale is given by 
$\langle \Theta \rangle /{M_{P}} \approx \lambda$, 
the mass matrix of up-quarks is given by 
\begin{equation}
M_{ij}=y_{ij} \langle H \rangle \lambda^{f_{i}+f_{j}+x}.
\end{equation}
The U(1)$_F$ charge $x$ of the Higgs field $H$ may fix 
the overall magnitude of the mass matrix. 
If one takes $x = -4$, the power of the superfield $\Theta$ 
becomes negative for $i=3$ or $j=3$, so that the corresponding 
elements in the mass matrix are prohibited and set 
to be zeros, which is called the SUSY zeros~\cite{SUSY}. 
The presence of zero entries in the mass matrix owing to 
the SUSY zeros leads to the twisting among $\five^*$ 
representations~\cite{Bando-Kugo}. 
Although it should be dynamically clarified whether 
this turning around mechanism of hierarchy is consistent 
with the Froggatt-Nielsen mechanism, so far we do not 
understand so well the dynamics of how and why the 
Froggatt-Nielsen mechanism works. 
The mechanism itself is, however, what we want to have, 
so that it seems to be inevitable in the study of mass 
matrices.

In these circumstances, we study the possibility of the 
twisting among the down-type quarks in the SUSY $E_6$ GUT 
without using the SUSY zeros. 
In the following, we restrict ourselves to the case of $x=0$ 
so that there is no entry in the mass matrix which is negative 
power of $\lambda$. 
Then we can ``naively'' use the Froggatt-Nielsen mechanism 
in order to generate the hierarchical structure of the mass 
matrix. 
We will diagonalize the $6\times 6$ mass matrix of the 
down-quarks which consists of 3 generations of 
$d_R^{c}$ in $(\textbf{16},\five^*)$ and $D_R^c$ in 
$(\ten,\five^*)$, and study if the appropriate quark masses 
and mixings could be obtained due to the twisting among them.  

Let us briefly review our scenario. 
We prepare 3 superfields, $\Psi_{i}(\textbf{27}) (i=1,2,3)$, 
corresponding to the 3 generations of quarks and leptons.  
We also introduce two Higgs fields, $H(\textbf{27})$ and 
$\phi(\textbf{78})$. 
In $\textbf{78}$, there are $\textbf{8}_{L}$ and $\textbf{8}_{R}$ 
under the decomposition of $E_6$ into $\mbox{SU(3)}_C \times 
\mbox{SU(3)}_L \times \mbox{SU(3)}_R$. 
We assume that the components of $\textbf{8}_L$ and 
$\textbf{8}_R$ develop the vacuum expectation values 
(VEVs) of the GUT scale ($\sim \mathcal{O}(10^{16}\gev)$) 
so that $E_6$ is broken down to 
$\mbox{SU(3)}_C \times \mbox{SU(2)}_L \times \mbox{U(1)}_Y 
\times \mbox{U(1)}_{\psi} \times \mbox{U(1)}_{\chi}$ 
directly. 
Then, in general, the baryon number violating operators in the 
superpotential are allowed from the gauge invariance and the 
proton stability cannot be preserved. 
We, therefore, introduce a discrete symmetry which is assigned 
to be odd for $\Psi_{i}(\textbf{27})$ and even for 
$H(\textbf{27})$~\cite{Bando-Kugo}. 
Another source of the proton decay is some Higgs fields 
in $\five^*$ and $\ten$, which carry both the weak and 
color charges, and they are assumed to be sufficiently heavy 
owing to some unknown mechanisms. 

There are two extra U(1) symmetries besides the SM 
gauge symmetry. 
As is studied in the superstring inspired $E_6$ model, 
the breaking scale of the extra U(1) symmetries could 
be lowered to $\mathcal{O}(\tev)$ without conflicting 
the electroweak precision measurements at LEP1 and 
SLC~\cite{Cho}. 
We assume that the component fields of $H(\textbf{27})$, 
which are singlets under the SM gauge group, play role 
to break the extra U(1) symmetries, \ie, 
we suppose that the VEV $\langle H(\textbf{16},\textbf{1}) 
\rangle$ breaks $\mbox{U(1)}_{\psi} \times \mbox{U(1)}_{\chi}$ 
down to ${\rm U(1)}'$ and the VEV 
$\langle H(\textbf{1},\textbf{1}) \rangle$ breaks 
${\rm U(1)}'$. 
These VEVs may appear in the mass matrices of quarks 
and leptons, if their scale is around $\mathcal{O}(\tev)$, 
which may be attainable in the future colliders, 

In practice, we examine the mass and mixing matrices of 
down-type quarks employing the perturbation theory. 
Although it is rather hard to calculate the realistic 
mass and mixing values for quarks based on the 
perturbative treatment, it is worth to use this method 
to clarify the characteristic features of the mass and mixing 
matrices in our model. 
We show that, in a certain parameter set, the hierarchical 
structure of the CKM matrix can be found due to the twisting. 
As a by-product, there is a light vector-like down-quark whose 
mass is around $\mathcal{O}(\tev)$. 

The twisting which we found also lead to the large mixing 
in the leptonic sector and the result will be reported 
elsewhere~\cite{obara}. 
It is worth to mention that the 6 $\times$ 6 down-quark mass 
matrix has been studied in the superstring inspired 
$\mbox{SU(6)} \times \mbox{SU(2)}_R$ model 
(Gepner model)~\cite{Matsu} 
in which some elements of the mass matrix vanish from 
the gauge symmetry.

This paper is organized as follows: the mass and mixing matrices 
of quarks in our model are examined in Sec.~\ref{section:framework}. 
The quantitative estimations of mass and mixing matrices will 
be done in Sec.~\ref{section:estimation}. 
Sec.~\ref{section:summary} is devoted to summary and discussion. 
Some formulae used in the calculation based on the perturbative 
method are given in detail in Appendix. 

\section{Mass and Flavor Mixing Matrices}
\label{section:framework}
\subsection{Superpotential}
\clean

We first review the superpotentials in SUSY $E_6$ GUT 
following ref.~\cite{Bando-Kugo}. 
As stated in the previous section, we introduce the flavor 
symmetry U(1)$_F$ and assume that the Froggatt-Nielsen 
mechanism~\cite{Frog} works on. 
Then, the superpotential relevant to the Higgs superfield 
$H(\textbf{27})$ is given by 
\begin{eqnarray}
W_{H} = y_{ij} \Psi_{i}(\textbf{27}) 
     \Psi_{j}(\textbf{27}) H(\textbf{27}) 
\left( \frac{\Theta}{M_{P}} \right)^{f_i + f_j}, 
\label{wh}
\end{eqnarray}
where $f_i$ are the U(1)$_F$ charges of the $i$-th generation 
quarks and leptons, and the Yukawa coupling constant $y_{ij}$ 
is assumed to be of the order one. 

The superpotential relevant to the adjoint Higgs superfield 
$\phi(\textbf{78})$ is given by the higher dimensional 
operators: 
\begin{eqnarray}
W_{\phi} = \sum_{i,j} s_{ij} M_{P}^{-1} \Psi_{i}(\textbf{27}) 
           \Psi_{j}(\textbf{27}) ( \phi(\textbf{78}) 
           H(\textbf{27}) )_{\bf{27}} 
           \left( \frac{\Theta}{M_{P}} \right)^{f_i + f_j} 
\nonumber \\ 
         + \sum_{i,j} a_{ij} M_{P}^{-1}
           ( \phi(\textbf{78}) \Psi_{i}(\textbf{27}) )_{\bf{27}}
           \Psi_{j}(\textbf{27}) H(\textbf{27})
           \left( \frac{\Theta}{M_{P}} \right)^{f_i + f_j}, 
\label{superpotential}
\end{eqnarray}
where $s_{ij}$ and $a_{ij}$ are symmetric and anti-symmetric 
tensors with respect to the generation indices, respectively. 
Here we take the U(1)$_F$ charges of Higgs fields to be zero 
according to the discussion in the previous section. 
The coupling constants $s_{ij}$ and $a_{ij}$ are also assumed 
to be of the order one. 
In eq.~(\ref{superpotential}), two Higgs fields are multiplied 
such as $(\phi(\textbf{78}) H(\textbf{27}) )_{\bf{27}}$, 
which denotes the infinitesimal 
transformation of the fundamental representation $\textbf{27}$ 
by the adjoint representation $\textbf{78}$. 
Therefore, $M_{P}^{-1}$ has to be introduced to modify the 
dimensionality.

The adjoint Higgs $\phi(\textbf{78})$ can be decomposed under 
the subgroup $\mbox{SU(3)}_{L} \times \mbox{SU(3)}_{R} 
\times \mbox{SU(3)}_{C} \subset E_6$ as follows: 
\begin{eqnarray}
\textbf{78} = \textbf{8}_{L} + \textbf{8}_{R} + \textbf{8}_{C}
            + (\textbf{3},\textbf{3},\textbf{3})
            + (\textbf{3}^*,\textbf{3}^*,\textbf{3}^*). 
\end{eqnarray}
Suppose that the components of $\textbf{8}_{R}$ and $\textbf{8}_{L}$ 
develop the VEVs as follows: 
\bsub
\label{VEV for adjoint Higgs}
\begin{eqnarray}
\frac{\langle \phi (\textbf{8}_{R}) \rangle}{M_{P}} 
&=& \left(
       \begin{array}{@{\,}ccc@{\,}}
       \omega + \chi_{R} & 0   & 0 \\
       0       & - \omega + \chi_{R} & 0 \\
       0       & 0     & - 2 \chi_{R}
       \end{array}
    \right), \\
\frac{\langle \phi (\textbf{8}_{L}) \rangle}{M_{P}} 
&=& \left(
       \begin{array}{@{\,}ccc@{\,}}
       \chi_{L} & 0    & 0 \\
       0       & \chi_{L} & 0 \\
       0       & 0     & - 2 \chi_{L}
       \end{array}
    \right),
\end{eqnarray}
\label{eq:adjoint-vev}
\esub
then $E_6$ is broken down to $\mbox{SU(3)}_C \times 
\mbox{SU(2)}_L \times \mbox{U(1)}_Y \times 
\mbox{U(1)}_{\psi} \times \mbox{U(1)}_{\chi}$. 
As a consequence, the mass matrices for the down-quarks and 
charged-leptons can be different. 
Even if the breaking scale $\langle \phi(\textbf{78}) \rangle$ is 
the GUT scale, $\sim \mathcal{O}(10^{16}\gev)$, the modification 
for the Yukawa couplings in eq.~(\ref{superpotential}) is 
expressed by the ratio $\langle \phi(\textbf{78}) \rangle /M_{P}$ 
so that it is reasonably small. 
\subsection{Up-quark sector}

Owing to the U(1)$_F$ charge assignment (\ref{eq:u1charge}) 
and the superpotentials (\ref{wh}) and (\ref{superpotential}), 
the Yukawa matrix for the up-quark sector has the hierarchical 
structure which is phenomenologically acceptable: 
\begin{eqnarray}
Y_{u} \equiv \left(
\begin{array}{@{\,}ccc@{\,}}
Y_{11} \lambda^6 & Y_{12} \lambda^5 & Y_{13} \lambda^3 \\ 
Y_{21} \lambda^5 & Y_{22} \lambda^4 & Y_{23} \lambda^2 \\ 
Y_{31} \lambda^3 & Y_{32} \lambda^2 & Y_{33} \end{array}
\right),
\end{eqnarray}
where
\begin{eqnarray}
Y_{ij} = y_{ij} + ( \chi_{R} - \chi_{L} + \omega ) s_{ij} 
       + \frac{1}{2} ( \chi_{R} + \chi_{L} + \omega ) a_{ij}. 
\end{eqnarray}
The Higgs field, $H(\textbf{27})$, is decomposed following 
eq. (\ref{eq:decomposition}). 
Then, $H(\ten,\five)$ couples to the up-type quarks and 
its VEV gives the mass matrix as follows: 
\begin{eqnarray}
M_{u} \equiv \left(
\begin{array}{@{\,}ccc@{\,}}
Y_{11} \lambda^6 & Y_{12} \lambda^5 & Y_{13} \lambda^3 \\ 
Y_{21} \lambda^5 & Y_{22} \lambda^4 & Y_{23} \lambda^2 \\ 
Y_{31} \lambda^3 & Y_{32} \lambda^2 & Y_{33} \end{array}
\right)
v \sin \beta,
\end{eqnarray}
where $\langle H(\textbf{10},\textbf{5})\rangle = 
v\sin\beta \equiv v_u$ is the VEV of $H(\textbf{10},\textbf{5})$. 
The VEV $v_u$ is normalized as $v^2 \equiv v_{u}^{2} + v_{d}^{2}
= (v \sin\beta)^2 + (v \cos\beta)^2 \simeq (174\gev)^2$, 
where $v\cos\beta\equiv v_d$ provides the mass to the 
down-type quarks. 
The mass matrix $M_{u}$ can be diagonalized by using the 
unitary matrix $U_{u_{L}}$ as follows: 
\begin{eqnarray}
U_{u_{L}}(M_{u}M_{u}^{\dagger})U_{u_{L}}^{\dagger} = 
{\rm diag}(m_{u}^2,m_{c}^2,m_{t}^2), 
\end{eqnarray}
where the mass eigenvalues are given by 
\bsub
\begin{eqnarray}
m_{u}^2 &=& f_{u}^2\, v^2 \sin^2 \beta , \\ 
m_{c}^2 &=& f_{c}^2\, v^2 \sin^2 \beta , \\ 
m_{t}^2 &=& f_{t}^2\, v^2 \sin^2 \beta . 
\end{eqnarray}
\label{eqs:up-quarks}
\esub
In eq. (\ref{eqs:up-quarks}), $f_u^2$, $f_c^2$ and $f_t^2$ 
represent the eigenvalues for the Yukawa matrix, whose 
magnitudes are given as follows \cite{Bando-Kugo}: 
\bsub
\bea
f_u^2 &\sim& \mathcal{O}((\lambda^6)^2), 
\\
f_c^2 &\sim & \mathcal{O}((\lambda^4)^2), 
\label{eq:yukawa-charm}
\\
f_t^2 &\sim& \mathcal{O}(1). 
\eea
\esub
\subsection{Down-quark sector}
\label{subsection:down}

We study the mass matrix for the down-quark sector, having
totally 6 flavors, \ie, $d_i$ and $D_i (i=1,2,3)$. 
Then, the $6\times 6$ Yukawa matrix for the down-quarks is 
expressed by the four blocks of the 3 $\times$ 3 matrices, 
\begin{eqnarray}
Y_{d} \equiv \left(
\begin{array}{@{\,}cc@{\,}}
Y_{u} + \alpha \tilde{s} & Y_{u} + \epsilon \tilde{s} \\ 
Y_{u} + \alpha \tilde{s} + \gamma \tilde{s}^{T} & 
       Y_{u} + \epsilon \tilde{s} + \gamma \tilde{s}^{T} 
\end{array}
\right),
\end{eqnarray}
where we define 
\bsub
\begin{eqnarray}
\label{alp}
\alpha &\equiv& -2 \omega , \\
\epsilon &\equiv& -(\omega + 3 \chi_{R}) , \\
\label{gam}
\gamma &\equiv& 3 \chi_{L} , \\
\tilde{s} &\equiv& \left(s_{ij} + \frac{1}{2} a_{ij}
\right) \lambda^{f_i + f_j}.
\end{eqnarray}
\esub
As shown in eq. (\ref{eq:adjoint-vev}), their order is 
$\mathcal{O}(\phi(\textbf{78})/M_P)$. 
Parametrizing the VEVs of the Higgs fields which couple to 
$d_i$ and $D_i$ as 
\bsub
\bea
\langle H(\textbf{10},\textbf{5}^*) \rangle 
    &=& v_{d} \cos\theta, \\
\langle H(\textbf{16},\textbf{5}^*) \rangle 
    &=& v_{d} \sin\theta, \\
\langle H(\textbf{16},\textbf{1}) \rangle 
     &=& v_{\!D \atop {}}^{ }\! \cos\varphi, \\
\langle H(\textbf{1},\textbf{1}) \rangle 
     &=& v_{\!D \atop {}}^{ }\! \sin\varphi, 
\eea
\esub
the mass matrix for the down-quarks can be expressed as 
follows:
\begin{eqnarray}
\label{Md}
M_{d} &=& \left(
\begin{array}{@{\,}cc@{\,}}
(Y_{u} + \alpha \tilde{s}) v_{d} \cos \theta & 
(Y_{u} + \epsilon \tilde{s}) v_{d} \sin \theta \\ 
(Y_{u} + \alpha \tilde{s} + \gamma \tilde{s}^{T})
v_{\!D \atop {}}^{ }\! \cos \varphi & 
(Y_{u} + \epsilon \tilde{s} + \gamma \tilde{s}^{T}) 
v_{\!D \atop {}}^{ }\! \sin \varphi 
\end{array}
\right).
\end{eqnarray}
It is easy to see that each block in eq.~(\ref{Md}) 
includes the Yukawa matrix of the up-quarks. 
Let us take $\alpha$, $\epsilon$ and $\gamma$ as 
the parameters of perturbation by assuming 
\bea
\alpha, \epsilon, \gamma \sim  
\mathcal{O}(\langle \phi(\textbf{78}) \rangle /M_{P}) 
\sim \lambda^4. 
\label{eq:epsilon}
\eea
Then we diagonalize the mass matrix using the following 
decomposition:
\begin{eqnarray}
M_{d} \equiv M^{(0)} + M^{(1)} ,
\label{decom}
\end{eqnarray}
where $M^{(0)}$ is the unperturbed mass matrix and $M^{(1)}$ is 
its perturbation. 
Their explicit forms are given by 
\bsub
\begin{eqnarray}
M^{(0)} &=& Y_{u} \otimes
\left(
\begin{array}{@{\,}cc@{\,}}
v_{d} \cos \theta & v_{d} \sin \theta \\ 
v_{\!D \atop {}}^{ }\! \cos \varphi & 
v_{\!D \atop {}}^{ }\! \sin \varphi 
\end{array}
\right) ,\\
M^{(1)} &=& \tilde{s} \otimes
\left(
\begin{array}{@{\,}cc@{\,}}
\alpha v_{d} \cos \theta & \epsilon v_{d} \sin \theta \\ 
\alpha v_{\!D \atop {}}^{ }\! \cos \varphi & 
\epsilon v_{\!D \atop {}}^{ }\! \sin \varphi 
\end{array}
\right) \! + \!
\tilde{s}^{T} \otimes
\left(
\begin{array}{@{\,}cc@{\,}}
0 & 0 \\
\gamma v_{\!D \atop {}}^{ }\! \cos \varphi & 
\gamma v_{\!D \atop {}}^{ }\! \sin \varphi 
\end{array}
\right) \! .
\end{eqnarray}
\label{eq:purturbation}
\esub

The mass matrix $M_d$ is diagonalized by using 
the unitary matrix $U_{d_{L}}$ as 
\begin{eqnarray}
U_{d_{L}}(M_{d}M_{d}^{\dagger})U_{d_{L}}^{\dagger} =
{\rm diag}(m_{d_1}^2,m_{d_2}^2,m_{d_3}^3,m_{D_1}^2, 
m_{D_2}^2, m_{D_3}^2), 
\end{eqnarray}
where $M_d M_d^\dagger$ is expressed by using (\ref{decom}) as 
\bsub
\begin{eqnarray}
M_{d}M_{d}^{\dagger}&=& M^{(0)} M^{(0)\dagger} 
             + \delta\mathcal{M}^2, 
\\
\delta\mathcal{M}^2 &\equiv& M^{(0)} M^{(1)\dagger} 
             + M^{(1)} M^{(0)\dagger} + M^{(1)} M^{(1)\dagger} .
\end{eqnarray}
\esub
Correspondingly, the unitary matrix $U_{d_{L}}$ is expanded as 
\begin{eqnarray}
U_{d_{L}} &\equiv& U_{d_{L}}^{(0)} + U_{d_{L}}^{(1)}.
\end{eqnarray}
The mass eigenvalues at the leading order are given by 
\begin{eqnarray}
U_{d_{L}}^{(0)}(M^{(0)} M^{(0)\dagger})U_{d_{L}}^{(0)\dagger} 
\!&=&
{\rm diag}\!\left(
(m_{d_1}^2)^{\!\mathstrut^{(\!0\!)}}, \cdots, 
(m_{d_3}^2)^{\!\mathstrut^{(\!0\!)}}, 
(m_{D_1}^2)^{\!\mathstrut^{(\!0\!)}}, \cdots, 
(m_{D_3}^2)^{\!\mathstrut^{(\!0\!)}}
\right)\!\!. 
\label{leading_mass}
\end{eqnarray}
The correction of the mass eigenvalue 
in the lowest order of the perturbation 
$\Delta_{n}^{(1)}$ is given by
\begin{eqnarray}
\Delta_{n}^{(1)} 
        =\langle n^{(0)} | \delta\mathcal{M}^2 | n^{(0)} \rangle ,
\end{eqnarray}
and the lowest order correction for the mass eigenstate 
$| n^{(1)} \rangle$ is given by
\begin{eqnarray}
\label{eigens}
| n^{(1)} \rangle =
\sum_{n \neq k} \frac{| k^{(0)} \rangle
\langle k^{(0)} | \delta\mathcal{M}^2 | n^{(0)} \rangle} 
{(m_n^2)^{\mathstrut^{(\!0\!)}}  - (m_k^2)^{\mathstrut^{(\!0\!)}} }, 
\end{eqnarray}
where $n,k= d_1,d_2,d_3,D_1,D_2,D_3$. 
The generic form of 
$\langle k^{(0)} | \delta\mathcal{M}^2 | n^{(0)} \rangle$ 
can be found in Appendix. 

If we assume $v_{d} \ll v_{\!D \atop {}}^{ }$ as a natural 
consequence of our scenario, we find the mass eigenvalues 
(\ref{leading_mass}) and the unitary matrix $U_{d_L}^{(0)}$ 
at the leading order as follows:
\bsub
\bea
U_{d_{L}}^{(0)}(M^{(0)} M^{(0)\dagger})U_{d_{L}}^{(0)\dagger} 
&=& \left(
             \begin{array}{@{\,}ccc@{\,}}
               f_{u}^2 & 0     & 0 \\
               0     & f_{c}^2 & 0 \\
               0     & 0     & f_{t}^2
             \end{array}
\right)
\otimes \left(
             \begin{array}{@{\,}cc@{\,}}
               v_d^2 \sin^2 (\theta - \varphi) & 0 \\
               0     & v_{\!D \atop {}}^2
             \end{array}
        \right) ,
\label{m-leading}
\\
U_{d_{L}}^{(0)} &=&  
U_{u_{L}} \otimes
\left(
\begin{array}{@{\,}cc@{\,}}
1 & - \frac{\mathstrut v_{d}}{v_{\!D \atop { }}} 
\cos (\theta - \varphi) \\ 
\frac{\mathstrut v_{d}}{v_{\!D \atop { }}} \cos (\theta - \varphi) & 1 
\end{array}
\right).
\label{u-d0}
\eea
\esub
Then the mass eigenstates for leading order $| n^{(0)} \rangle$'s
correspond to the six columns of the matrix $U_{d_{L}}^{(0)\dagger}$.
So, at the lowest order of the perturbation, the mass eigenvalues 
are given as follows: 
\bsub
\begin{eqnarray}
\label{md1}
m_{d_{1}}^2 &\simeq& \frac{m_{u}^2}{\tan^2 \beta} \sin^2 (\theta -
\varphi) 
+ \Delta_{d_1}^{(1)},\\
m_{d_{2}}^2 &\simeq& \frac{m_{c}^2}{\tan^2 \beta} \sin^2 (\theta -
\varphi) 
+ \Delta_{d_2}^{(1)},\\
m_{d_{3}}^2 &\simeq& \frac{m_{t}^2}{\tan^2 \beta} \sin^2 (\theta -
\varphi) 
+ \Delta_{d_3}^{(1)},\\
m_{D_{1}}^2 &\simeq& v_{\!D \atop { }}^2f_{u}^2 + \Delta_{D_1}^{(1)},\\ 
m_{D_{2}}^2 &\simeq& v_{\!D \atop { }}^2f_{c}^2 + \Delta_{D_2}^{(1)},\\ 
\label{mD3}
m_{D_{3}}^2 &\simeq& v_{\!D \atop { }}^2f_{t}^2 + \Delta_{D_3}^{(1)}. 
\end{eqnarray}
\esub
On the other hand the correction of the mass eigenstate 
for the lowest order 
$| n^{(1)} \rangle$ corresponds to each column of the 
unitary matrix $U_{d_{L}}^{(1)\dagger}$.

From eq.~(\ref{m-leading}), it is clear that 
$m_{d_1}<m_{d_2}<m_{d_3}$ and $m_{D_1}<m_{D_2}<m_{D_3}$, 
but $m_{d_3}<m_{D_1}$ may not hold in general. 
Even under the assumption $v_d \ll v_D$, for example, 
$m_{d_3}$ could be heavier than $m_{D_1}$ and $m_{D_2}$ 
in a certain parameter space. 
But this possibility is unacceptable. 
Let us recall that the CKM matrix has the small off-diagonal 
elements. 
In our model, it could be explained by the structure of 
the unitary matrix $U_{d_L}$. 
Since the leading order of the matrix $U_{d_L}$ (\ref{u-d0}) 
is proportional to the matrix $U_{u_L}$, the CKM matrix is given 
as the unit matrix at the leading order only if the (1,1) block 
of r.h.s. in eq. (\ref{u-d0}) is identified as the light 
down-quarks. The non-vanishing off-diagonal elements of the CKM matrix 
are derived from the lowest order of the perturbation, $U_{d_L}^{(1)}$, 
so that they become small.
One may consider the (2,2) block of r.h.s. in eq. (\ref{u-d0}) 
is also another candidate of the light down-quarks, however, 
the condition $v_d \ll v_D$ does not allow this possibility. 
We, therefore, identify $d_1,d_2$ and $d_3$ as the ordinary 
light down-quarks, $d,s$ and $b$. 

Now the ordinary $3\times 3$ CKM matrix can be defined by 
\begin{eqnarray}
\label{CKM123}
V_{i j} &\equiv& \sum_{\alpha=1}^3 
(U_{u_{L}})_{i \alpha}(U_{d_{L}}^{\dagger})_{\alpha j} ,
\end{eqnarray}
where the indices $i$ and $j$ denote $u,c,t$ and $d,s,b$, 
respectively. 
Using new symbols $m_+^2 \equiv v_D^2$ and 
$m_-^2 \equiv v_d^2 \sin^2(\theta - \varphi)$, 
we explicitly write down the CKM matrix elements as follows: 
\bsub
\begin{eqnarray}
\label{Vud}
V_{u d} &\simeq& 1 + \frac{v_{d}}
{v_{\!D \atop { }}} \cos (\theta - \varphi) 
\frac{\langle D_1 | \delta\mathcal{M}^2 | d_1 \rangle}
{ f_{u}^2 (m_{-}^2 - m_{+}^2)} , 
\\ 
V_{u s} &\simeq& \frac{\langle d_1 | \delta\mathcal{M}^2 | 
     d_2 \rangle} {m_{-}^2 (f_{c}^2 - f_{u}^2)} 
+ \frac{v_{d}}{v_{\!D \atop { }}} \cos (\theta - \varphi) 
\frac{\langle D_1 | \delta\mathcal{M}^2 | d_2 \rangle}
{ m_{-}^2 f_{c}^2 - m_{+}^2 f_{u}^2} , 
\\ 
V_{u b} &\simeq& \frac{\langle d_1 | \delta\mathcal{M}^2 | 
     d_3 \rangle}{m_{-}^2 (f_{t}^2 - f_{u}^2)} 
+ \frac{v_{d}}{v_{\!D \atop { }}} \cos (\theta - \varphi) 
\frac{\langle D_1 | \delta\mathcal{M}^2 | d_3 \rangle}
{ m_{-}^2 f_{t}^2 - m_{+}^2 f_{u}^2} , 
\\ 
V_{c d} &\simeq& \frac{\langle d_2 | \delta\mathcal{M}^2 | 
    d_1 \rangle}{m_{-}^2 (f_{u}^2 - f_{c}^2)} 
+ \frac{v_{d}}{v_{\!D \atop { }}} \cos (\theta - \varphi) 
\frac{\langle D_2 | \delta\mathcal{M}^2 | d_1 \rangle}
{ m_{-}^2 f_{u}^2 - m_{+}^2 f_{c}^2} , 
\\ 
V_{c s} &\simeq& 1 + \frac{v_{d}}
{v_{\!D \atop { }}} \cos (\theta - \varphi) 
\frac{\langle D_2 | \delta\mathcal{M}^2 | d_2 \rangle}
{ f_{c}^2 (m_{-}^2 - m_{+}^2)} , 
\\ 
V_{c b} &\simeq& \frac{\langle d_2 | \delta\mathcal{M}^2 | d_3 \rangle}
{m_{-}^2 (f_{t}^2 - f_{c}^2)} 
+ \frac{v_{d}}{v_{\!D \atop { }}} \cos (\theta - \varphi) 
\frac{\langle D_2 | \delta\mathcal{M}^2 | d_3 \rangle}
{ m_{-}^2 f_{t}^2 - m_{+}^2 f_{c}^2} , 
\\ 
V_{t d} &\simeq& \frac{\langle d_3 | \delta\mathcal{M}^2 | d_1 \rangle}
{m_{-}^2 (f_{u}^2 - f_{t}^2)} 
+ \frac{v_{d}}{v_{\!D \atop { }}} \cos (\theta - \varphi) 
\frac{\langle D_3 | \delta\mathcal{M}^2 | d_1 \rangle}
{ m_{-}^2 f_{u}^2 - m_{+}^2 f_{t}^2} , 
\\ 
V_{t s} &\simeq& \frac{\langle d_3 | \delta\mathcal{M}^2 | d_2 \rangle}
{m_{-}^2 (f_{c}^2 - f_{t}^2)} 
+ \frac{v_{d}}{v_{\!D \atop { }}} \cos (\theta - \varphi) 
\frac{\langle D_3 | \delta\mathcal{M}^2 | d_2 \rangle}
{ m_{-}^2 f_{c}^2 - m_{+}^2 f_{t}^2} , 
\\ 
\label{Vtb}
V_{t b} &\simeq& 1 + \frac{v_{d}}
{v_{\!D \atop { }}} \cos (\theta - \varphi) 
\frac{\langle D_3 | \delta\mathcal{M}^2 | d_3 \rangle}
{ f_{t}^2 (m_{-}^2 - m_{+}^2)} . 
\end{eqnarray}
\label{eq:ckm-expression}
\esub
\section{Estimation of Masses and Mixings}
\label{section:estimation}
\clean

In this section, we estimate the mass and mixing matrices 
quantitatively in a certain set of the parameters. 
In the case of $\alpha = \gamma = 0$ and $\epsilon \neq0$, 
that is, $\chi_L=\omega=0$ and $\chi_R \neq0$, 
the mass eigenvalues eqs.~(\ref{md1}) $\sim$ (\ref{mD3}) 
are given as follows: 
\bsub
\begin{eqnarray}
m_{d}^2 &\sim& \frac{m_{u}^2}{\tan^2 \beta} 
\sin^2 (\theta - \varphi)
+ \epsilon \,v_{d}^2 \, \lambda^6 \,
( \sin\theta - \cos(\theta - \varphi) )^2,\\
m_{s}^2 &\sim& \frac{m_{c}^2}{\tan^2 \beta} 
\sin^2 (\theta - \varphi)
+ \epsilon \,v_{d}^2 \, \lambda^4 \,
( \sin\theta - \cos(\theta - \varphi) )^2,\\
m_{b}^2 &\sim& \frac{m_{t}^2}{\tan^2 \beta} 
\sin^2 (\theta - \varphi)
+ \epsilon \,v_{d}^2 \,
( \sin\theta - \cos(\theta - \varphi) )^2, 
\label{eq-bottom}
\\
m_{D_{1}}^2 &\sim& v_{\!D \atop { }}^2 f_{u}^2
+ \epsilon \,v_{\!D \atop { }}^2 \lambda^6 \sin^2 \varphi,\\ 
m_{D_{2}}^2 &\sim& v_{\!D \atop { }}^2 f_{c}^2
+ \epsilon \,v_{\!D \atop { }}^2 \lambda^4 \sin^2 \varphi,\\ 
m_{D_{3}}^2 &\sim& v_{\!D \atop { }}^2 f_{t}^2 
+ \epsilon \,v_{\!D \atop { }}^2 \sin^2 \varphi, 
\end{eqnarray}
\esub
and the CKM matrix elements eqs. (\ref{Vud}) $\sim$ (\ref{Vtb}) 
are given by 
\bsub
\begin{eqnarray}
V_{ud} &\sim& 1 , \\
V_{us} &\sim& \frac{ - \epsilon \lambda^5 ( \sin\theta - \cos(\theta -
\varphi)\sin\varphi )^2 } {\sin^2(\theta - \varphi)
f_{c}^2} , 
\label{eq:vus}
\\ 
V_{ub} &\sim& \frac{ - \epsilon \lambda^3 ( \sin\theta -
\cos(\theta - \varphi)\sin\varphi )^2 } {\sin^2(\theta -
\varphi) f_{t}^2} , 
\label{eq:vub}
\\ 
V_{cd} &\sim& - V_{us} , \\
V_{cs} &\sim& 1 , \\
V_{cb} &\sim& \frac{ - \epsilon \lambda^2 ( \sin\theta - \cos(\theta -
\varphi)\sin\varphi )^2 } {\sin^2(\theta - \varphi)
f_{t}^2} , 
\label{eq:vcb}
\\ 
V_{td} &\sim& - V_{ub} , \\
V_{ts} &\sim& - V_{cb} , \\
V_{tb} &\sim& 1 .
\end{eqnarray}
\label{eq:vckm}
\esub
We examine the masses and mixing angles at the GUT scale 
with the following inputs: 
\renewcommand{\theenumi}{\roman{enumi}}
\renewcommand{\labelenumi}{(\theenumi)}
\begin{enumerate}
\item
$v_D \sim 10^4\tev$ so that the heavy down-quarks ($D_i$) 
do not decouple from the mass matrix,
\item
$\sin^2(\theta - \varphi) / \tan^2 \beta \sim (1/100)^2$ 
and $\tan\beta=40$ to reproduce the bottom quark mass 
($m_b \sim 1 \rm GeV$) in eq. (\ref{eq-bottom}),
\item
$\sin\theta \sim 0.4$, $\cos\theta \sim 0.9$, 
$\sin\varphi \sim \lambda^8$ and $\cos\varphi \sim 1$ 
so that $V_{us}$ in eq. (\ref{eq:vus}) is approximately 
equal to $\lambda$, taking account of 
eqs. (\ref{eq:yukawa-charm}) and (\ref{eq:epsilon}). 
\end{enumerate}
From our inputs (i) $\sim$ (iii), 
we find the mass eigenvalues of down-quarks as 
\bsub
\bea
m_d &\sim & 0.2~{\rm MeV}, 
\label{eq:downmass}
\\
m_s &\sim & 4~{\rm MeV}. 
\label{eq:strangemass}
\eea
\label{eq:d-s-quarks}
\esub
The obtained $d$- and $s$-quark masses (3.3) 
are a bit small compared with those in the MSSM 
at the GUT scale given in ref.~\cite{koide-fusaoka}. 
Then, the CKM matrix elements are obtained as 
\bea
V \sim \left(
\begin{array}{ccc}
1 & \lambda & \lambda^7 \\
\lambda & 1 & \lambda^6 \\
\lambda^7 & \lambda^6 &1 
\end{array}
\right). 
\eea
Therefore, the values for $V_{us}$ and $V_{cd} $ can be reasonably 
reproduced in our model, but the other off-diagonal elements 
of the CKM matrix are much smaller than those expected at the 
GUT scale~\cite{koide-fusaoka}. 
The smallness of $V_{ub}$ and $V_{cb}$ is caused 
by the hierarchical 
structure of the Yukawa couplings in eq. (2.10). 
(See (\ref{eq:vus}), (\ref{eq:vub}) and (\ref{eq:vcb}).) 

Next let us see $U_{d_{R}}$, which diagonalizes 
the mass matrix $M_d^\dagger M_d$ in the same manner 
as in Sec.~\ref{subsection:down}, 
in order to see that if the twisting among the down-type quarks 
occurs or not.  
After the tedious calculations, we find 
\bsub
\begin{eqnarray}
| d_{R} \rangle &=& \mathcal{O}(\lambda^6) | d'_{1_{R}} \rangle 
+ \mathcal{O}(\lambda^7) | d'_{2_{R}} \rangle
+ \mathcal{O}(\lambda^9) | d'_{3_{R}} \rangle \nonumber \\
&+& \underline{ \mathcal{O}(1) | D'_{1_{R}} \rangle }
+ \mathcal{O}(\lambda) | D'_{2_{R}} \rangle
+ \mathcal{O}(\lambda^3) | D'_{3_{R}} \rangle, 
\\[3mm]
| s_{R} \rangle &=& \mathcal{O}(\lambda^5) | d'_{1_{R}} \rangle 
+ \mathcal{O}(\lambda^6) | d'_{2_{R}} \rangle
+ \mathcal{O}(\lambda^8) | d'_{3_{R}} \rangle
\nonumber \\ 
&+& \mathcal{O}(\lambda) | D'_{1_{R}} \rangle
+ \underline{ \mathcal{O}(1) | D'_{2_{R}} \rangle }
+ \mathcal{O}(\lambda^2) | D'_{3_{R}} \rangle, 
\\[3mm]
| b_{R} \rangle &=& \mathcal{O}(\lambda^3) | d'_{1_{R}} \rangle 
+ \mathcal{O}(\lambda^4) | d'_{2_{R}} \rangle
+ \mathcal{O}(\lambda^6) | d'_{3_{R}} \rangle
\nonumber \\
&+& \mathcal{O}(\lambda^3) | D'_{1_{R}} \rangle
+ \mathcal{O}(\lambda^2) | D'_{2_{R}} \rangle
+ \underline{ \mathcal{O}(1) | D'_{3_{R}} \rangle }, 
\end{eqnarray}
\label{eq:twist}
\esub
where the l.h.s. and the r.h.s. denote the mass and the current 
eigenstates, respectively. 
The state with the underline in the r.h.s. is the dominant 
component in the mass eigenstate. 
From eq. (\ref{eq:twist}), we find that the mass eigenstates for 
the right-handed down-type quarks, 
$| d_{R} \rangle$, $| s_{R} \rangle$ and $| b_{R} \rangle$ are 
mainly dominated by the current eigenstates, 
$| D'_{1_{R}} \rangle$, $| D'_{2_{R}} \rangle$ and 
$| D'_{3_{R}} \rangle$ in $(\ten,\five^*)$, respectively. 
This means that the twisting occurs between 
$(\textbf{10},\textbf{5}^*)$ and $(\textbf{16},\textbf{5}^*)$.  
As a complement of the above estimation based on the perturbation, 
we examined the mass and mixing matrices numerically 
and confirmed that the twisting occurs in the same 
parameter space. 
\section{Summary}
\label{section:summary}
\clean

In this paper we have studied the mass and mixing matrices of 
the supersymmetric $E_6$ GUT model, in which $E_6$ is assumed to 
be broken down to $\mbox{SU(3)}_C \times \mbox{SU(2)}_L 
\times \mbox{U(1)}_Y \times \mbox{U(1)}_{\psi} \times 
\mbox{U(1)}_{\chi}$ by 
the VEVs of the \textbf{78} Higgs scalar at GUT scale, while 
the extra U(1) symmetries are assumed to be broken 
by the VEVs of the component fields of the $\textbf{27}$ 
Higgs scalar taking around the energy scale, 
$\mathcal{O}(10^2 \gev)$--$\mathcal{O}(10^4 \tev)$. 
Quarks and leptons belong to the fundamental representation 
$\textbf{27}$ which contains two 
$\textbf{5}^{*}$'s --- $(\textbf{16},\five^*)$ and $(\ten,\five^*)$. 
Then, we have 6 flavors of down-type quarks. 
By diagonalizing the 6 $\times$ 6 mass matrix, we find 
the parameter regions in which the twisting ``naturally" occurs, 
having reasonable values for the CKM matrix elements, 
$V_{ud},V_{cs},V_{tb} \sim 1$ and $V_{us},V_{cd} \sim \lambda$, 
and the reasonable values for the down-quark masses, 
$m_d \sim 0.2 \rm MeV$, $m_s \sim 4 \rm MeV$ and $m_b \sim 1\gev$.
As a by-product, one vector-like down-quark is produced 
at TeV scale with $v_D \sim 10^4\tev$. 
We obtain, however, the rather small values for 
$V_{ub}$ and $V_{cb}$ because of the hierarchical 
structure of the Yukawa couplings.
%

In the derivation we employed the perturbation theory, in which 
the VEVs of the \textbf{78} Higgs scalar, 
$\langle \phi(\textbf{78}) \rangle /M_{P}$, are taken as the 
perturbations to those of $\textbf{27}$ Higgs scalar, 
$\langle H(\textbf{27}) \rangle /v$. 
Since the $\five^*$ multiplets consist of the right-handed down 
quarks and the left-handed leptons, 
the natural twisting found in this paper in the right-handed 
down-quark sector leads to the natural twisting in the left-handed 
lepton sector. 
Further study may lead to the understanding of the large neutrino 
mixings, hopefully the double maximal ones~\cite{obara}. 
%
%
%
%
\section*{Acknowledgements}

The authors would like to thank M. Bando for discussions and 
encouragements. They are also grateful to N. Oshimo for fruitful 
discussions and comments. This work is supported in part by 
Grant-in-Aid for Scientific Research from the Ministry of 
Education, Culture, Sports, Science and Technology of Japan 
(No. 11640262). 
%
%
%
%
\section*{Appendix}
\label{section:appendix}
\renewcommand{\theequation}{{\rm A}\arabic{equation}}
\clean

In this Appendix, we present the generic form of 
$\langle k^{(0)} | \delta\mathcal{M}^2 | n^{(0)} \rangle$, 
which has appeared in eq. (\ref{eigens}).
Here, for later convenience, we rewrite 
$d_1$, $\cdots$, $d_3$, $D_1$, $\cdots$, $D_3$ as follows:
$d_{1} \rightarrow 1$, $\cdots$, $d_{3} \rightarrow 3$, 
$D_{1} \rightarrow 4$, $\cdots$, 
$D_{3} \rightarrow 6$.

The perturbation term is composed of six parts: 
\begin{eqnarray}
\delta\mathcal{M}^2 &=& 
\delta\mathcal{M}^2(\alpha,\alpha^2) + 
\delta\mathcal{M}^2(\epsilon,\epsilon^2) 
+ \delta\mathcal{M}^2(\gamma,\gamma^2) \nonumber \\
&+& \delta\mathcal{M}^2(\alpha \,\epsilon)
+ \delta\mathcal{M}^2(\epsilon \,\gamma)
+ \delta\mathcal{M}^2(\alpha \,\gamma) ,
\end{eqnarray}
where the arguments $\alpha,\epsilon$ and $\gamma$ are the parameters of 
the perturbation and defined in eqs. (\ref{alp}) $\sim$ (\ref{gam}).

In the case of six flavors, the generic form of 
$\langle k^{(0)} | \delta\mathcal{M}^2 | n^{(0)} \rangle$ is 
given as: \\
for $k,n=1,2,3$, 
\begin{eqnarray}
\label{V123}
\langle k^{(0)} | \delta\mathcal{M}^2 | n^{(0)} \rangle &=&
\alpha \,v_{d}^2 Z_{k n} I_{C}(\theta,\varphi)^2 
+ \alpha^2 v_{d}^2 S_{k n} I_{C}(\theta,\varphi)^2 \nonumber \\
&+& \epsilon \,v_{d}^2 Z_{k n} I_{S}(\theta,\varphi)^2 
+ \epsilon^2 v_{d}^2 S_{k n} I_{S}(\theta,\varphi)^2 \nonumber \\
&+& \gamma^2 v_{d}^2 S_{k n} \cos(\theta - \varphi)^2 , 
\end{eqnarray}
for $k,n=4,5,6$, 
\begin{eqnarray}
\label{V456}
\langle k^{(0)} | \delta\mathcal{M}^2 | n^{(0)} \rangle &=&
\alpha \,v_{\!D \atop { }}^2 Z_{k-3\;n-3} \cos^2\varphi 
+ \alpha^2 v_{\!D \atop { }}^2 S_{k-3\;n-3} \cos^2\varphi \nonumber \\ 
&+& \epsilon \,v_{\!D \atop { }}^2 Z_{k-3\;n-3} \sin^2\varphi 
+ \epsilon^2 v_{\!D \atop { }}^2 S_{k-3\;n-3} \sin^2\varphi \nonumber \\ 
&+& \gamma \,v_{\!D \atop { }}^2 T_{k-3\;n-3} 
+ \gamma^2 v_{\!D \atop { }}^2 S_{k-3\;n-3} , 
\end{eqnarray}
for $k=4,5,6,n=1,2,3$, 
\begin{eqnarray}
\langle k^{(0)} | \delta\mathcal{M}^2 | n^{(0)} \rangle &=&
\alpha \,v_{d} \,v_{\!D \atop { }}^{ } 
Z_{k-3\;n} I_{C}(\theta,\varphi) \cos\varphi
+ \alpha^2 v_{d} \,v_{\!D \atop { }}^{ } 
S_{k-3\;n} I_{C}(\theta,\varphi) 
\cos\varphi \nonumber \\
&+& \epsilon \,v_{d} \,v_{\!D \atop { }}^{ } 
Z_{k-3\;n} I_{S}(\theta,\varphi) 
\sin\varphi
+ \epsilon^2 v_{d} \,v_{\!D \atop { }}^{ } 
S_{k-3\;n} I_{S}(\theta,\varphi) \sin\varphi \nonumber \\
&-& \gamma \,v_{d} \,v_{\!D \atop { }}^{ } 
Q_{k-3\;n} \cos(\theta - \varphi) 
- \gamma^2 v_{d} \,v_{\!D \atop { }}^{ } 
S_{k-3\;n} \cos(\theta - \varphi) , 
\end{eqnarray}
and for $k=1,2,3,n=4,5,6$, 
\begin{eqnarray}
\langle k^{(0)} | \delta\mathcal{M}^2 | n^{(0)} \rangle &=&
\alpha \,v_{d} \,v_{\!D \atop { }}^{ } 
Z_{k\;n-3} I_{C}(\theta,\varphi) \cos\varphi
+ \alpha^2 v_{d} \,v_{\!D \atop { }}^{ } 
S_{k\;n-3} I_{C}(\theta,\varphi) \cos\varphi \nonumber \\
&+& \epsilon \,v_{d} \,v_{\!D \atop { }}^{ } 
Z_{k\;n-3} I_{S}(\theta,\varphi) \sin\varphi
+ \epsilon^2 v_{d} \,v_{\!D \atop { }}^{ } 
S_{k\;n-3} I_{S}(\theta,\varphi) \sin\varphi \nonumber \\
&-& \gamma \,v_{d} \,v_{\!D \atop { }}^{ } 
(Q_{n-3\;k})^{T} \cos(\theta - \varphi) 
- \gamma^2 v_{d} \,v_{\!D \atop { }}^{ } 
S_{k\;n-3} \cos(\theta - \varphi) . 
\end{eqnarray}
In the above expressions, we define the various functions and 
coefficients as
\begin{eqnarray}
I_C(\theta,\varphi)
&\equiv& 
\cos\theta - \cos(\theta - \varphi)\cos\varphi, 
\\ 
I_S(\theta,\varphi)
&\equiv& 
\sin\theta - \cos(\theta - \varphi)\sin\varphi, 
\end{eqnarray}
and
\begin{eqnarray}
Q_{k n} &\equiv& \mathbf{u}_{k} Y_{u}\tilde{s} \mathbf{u}_{n}^{T},\\ 
S_{k n} &\equiv& \mathbf{u}_{k} \tilde{s}^{T} \tilde{s} 
\mathbf{u}_{n}^{T} ,\\
T_{k n} &\equiv& \mathbf{u}_{k} (Y_{u}\tilde{s} + \tilde{s}^{T}Y_{u}^{T}) 
\mathbf{u}_{n}^{T} ,\\
Z_{k n} &\equiv& \mathbf{u}_{k} (Y_{u}\tilde{s}^{T} + \tilde{s}Y_{u}^{T}) 
\mathbf{u}_{n}^{T} , 
\end{eqnarray}
where we define the three vectors $\textbf{u}_{i}$'s $(i=1,2,3)$ 
as the three rows of the unitary matrix $U_{u_L}$. 
We mention that $S_{k n},T_{k n}$ and $Z_{k n}$ are symmetric 
matrix, but $Q_{k n}$ is not symmetric. 
The generic form of $\Delta_{n}^{(1)}$ ($n=1,2,3$) is the same 
as the case of $n=k$ in eq. (\ref{V123}),  and that of 
$\Delta_{n}^{(1)}$ ($n=4,5,6$) is the same as the case of 
$n=k$ in eq. (\ref{V456}). 
%
%
%
%
%
%

\end{document}